\documentclass[aps,prl,twocolumn,showpacs]{revtex4}

\usepackage{graphicx}

\usepackage{amsmath} \newcommand{\EqLabel}[1]{\label{#1}}

\begin{document}
\title{Kubo formula for finite size systems}

\author{Jinshan Wu and Mona Berciu}

\affiliation{ Department of Physics and Astronomy, University of
  British Columbia, Vancouver, BC, Canada, V6T~1Z1}

\begin{abstract}
We demonstrate that the proper calculation of the linear response for
finite-size systems can only be performed if the coupling to the
leads/baths is explicitly taken into consideration. We exemplify this
by obtaining a Kubo-type formula for heat transport in a finite-size
system coupled to two thermal baths, kept at different
temperatures. We show that the proper calculation results in a
well-behaved response, without the singular contributions from degenerate
states encountered when Kubo formulae for infinite-size systems are
inappropriately used for finite-size systems. 
\end{abstract}

 \date{\today}

\maketitle

The Liouville-von Neumann equation of motion for the density matrix
$\rho(t)$ of a closed system is ($\hbar=1$):
\begin{equation}
\frac{\partial \rho\left(t\right)}{\partial t} = L_H
\rho\left(t\right)= -i[H, \rho],  \EqLabel{EOM}
\end{equation}
where $H$ is the Hamiltonian describing the evolution of the
system. If $H = H_0 + V$, where $V$ is a static weak coupling
to an external field, one can use perturbation theory to find a solution
$\rho\left(t\right) = \rho_{0}+\delta\rho\left(t\right)$ near a state
$\rho_0$ of the unperturbed system, $L_{H_0}\rho_{0}=0$. Neglecting
$L_{V} \delta\rho\left(t\right)$ in Eq. (\ref{EOM}), we 
integrate the resulting equation for $\delta\rho\left(t\right)$ to
obtain the $t\rightarrow \infty$, steady-state solution:
\begin{equation}
\delta\rho= \int_{0}^{\infty}dt e^{L_{0}t-\eta t} L_V \rho_{0},
\EqLabel{LRSsolution}
\end{equation}
where $\eta\rightarrow 0^{+}$. If $\rho_{0}=\frac{1}{Z}e^{-\beta
H_{0}}$ describes the unperturbed
system in equilibrium  at temperature $k_{B}T=1/\beta$, then
this is known as the Kubo formula~\cite{Kubo, Luttinger}:
\begin{equation}
\delta\rho = -i\int_{0}^{\infty}dt e^{-\eta t}\left[V\left(-t\right),
\frac{e^{-\beta H_0}}{Z}\right], \EqLabel{Kubo0}
\end{equation}
where $V(t) = e^{{i} H_0 t}V e^{-{i} H_0 t}$. We can use the identity 
$[V(-t),e^{-\beta H_{0}}]= -ie^{-\beta H_{0}}\int_{0}^{\beta} d\tau
\dot{V}\left(-t-i\tau\right)$ 
to rewrite:
\begin{equation}
\delta\rho = -\int_{0}^{\infty}dt e^{-\eta
t}\int_{0}^{\beta}d\tau\rho_0\dot{V}\left(-t-i\tau\right). \EqLabel{Kubo2}
\end{equation}
In terms of the eigenbasis $H_0|n\rangle = \epsilon_n|n\rangle$, 
$ \langle m|\dot{V}\left(t\right)|n\rangle  =
i\left(\epsilon_{m}-\epsilon_{n}\right)
V_{mn}e^{i\left(\epsilon_{m}-\epsilon_{n}\right)t}$, leading to:
\begin{align}
\delta\rho = \sum_{\substack{m,n \\ \epsilon_{m}\neq\epsilon_{n}
}}\frac{e^{-\beta\epsilon_{m}}-
  e^{-\beta\epsilon_{n}}}{Z}\frac{V_{mn}}{\epsilon_{m} -
  \epsilon_{n}-i\eta}\left|m\rangle\langle n\right|.  \EqLabel{KuboI}
\end{align} 
Note that there is no contribution from states with
$\epsilon_n=\epsilon_m$, for which $\langle
m|\dot{V}\left(t\right)|n\rangle =0$.  This also follows directly from 
Eq. (\ref{Kubo0}); if we 
write $V=V_0+V_{\perp}$, where $V_0=\sum_{
  \epsilon_{m}=\epsilon_{n}}V_{mn}\left|m\left>\right<n\right|$
commutes with $H_0$, then $[V(-t), \rho_0] = [V_{\perp}(-t),
  \rho_0]$. The ``diagonal'' part $V_0$ of $V$ does not contribute to
$\delta \rho$, and consequently has no influence on the static response functions.  

The lack of contributions from eigenstates with
$\epsilon_n=\epsilon_m$ is, however, 
puzzling, because  the well-known Drude weight derived 
directly from the Kubo formula is~\cite{Heidrich}:
\begin{align}
\label{Drude}
D=\frac{\pi\beta}{L} \sum_{\substack{ m,n \\ \epsilon_{m}=\epsilon_{n}
}}\frac{e^{-\beta\epsilon_{m}}}{Z}
|\langle m|\hat{J}|n\rangle |^{2},   
\end{align} 
i.e. it has contributions only from these eigenstates.

To understand the reason for this difference, consider the derivation
of Eq. (\ref{Drude}) from Eq. (\ref{Kubo2}), for simplicity for a
one-dimensional chain described by $H_{0} = -t
\sum_{l}\left(c^{\dag}_{l}c_{l+1}+h.c.\right) +
V_{0}\sum_{l}n_{l}n_{l+1} $, where $n_{l}=c^{\dag}_{l}c_{l}$, plus a
static potential $V=\sum_{l} V_{l}n_{l}$ induced by a
homogeneous electric field $E=-\nabla V$. From the continuity
equation,  $\dot{V}(t)=\sum_{l}V_{l}\frac{d}{dt}n_l(t) =
-{1\over a}\sum_{l}V_{l}[J_{l+1}(t)-J_{l}(t)]$, where $J_l
=it\left(c^+_{l+1}c_{l}-c^{+}_{l}c_{l+1}\right)$ is the local current
operator. This can be changed to $-{1\over
  a}\sum_{l}[V_{l-1}J_{l}(t)-V_{l}J_{l}(t)] = \nabla V
\sum_{l}J_{l}(t) = -EJ(t)$, where $J(t)$ is the total current
operator. Using $\dot{V}(t)= -EJ(t)$ in Eq. (\ref{Kubo2}) gives
$\delta\rho = E \int_{0}^{\infty}dt e^{-\eta
  t}\int_{0}^{\beta}d\tau\rho_0J\left(-t-i\tau\right)$. The dc
conductivity is then $ \sigma = \int_{0}^{\infty}dt e^{-\eta
  t}\int_{0}^{\beta} d\tau\left<J\left(-t-i\tau\right)J\right>$, where
$\langle O \rangle = Tr [\rho_0 O]$, from which Eq.  (\ref{Drude})
follows.

The only questionable step in this derivation, and the one responsible
for going from a result with no contributions from states with
$\epsilon_n=\epsilon_m$ to one with contributions only from these
states, is the change $\sum_{l}V_{l}J_{l+1}(t)\rightarrow
\sum_{l}V_{l-1}J_{l}(t)$. This is only justified for an
infinite system (where boundary terms are negligible), or a system
with periodic boundary conditions {\em and} an external electric
potential with the same periodicity~\cite{com1}. It is certainly not
valid for a 
finite size system connected to external leads, which break this
symmetry. In such cases,  the use of formulae like Eq. (\ref{Drude})
is simply not  
appropriate~\cite{note1}.

The solution, however, is not the use  of Eqs. (\ref{Kubo2}) or
(\ref{KuboI}),  derived
for a finite size, closed system. Instead, one needs 
to {\em derive their analogue for an open system coupled to leads}. The
reason is that $\tilde{\rho} = \rho_0 +
\delta \rho$ of Eq. (\ref{KuboI}) does not describe
a non-equilibrium stationary state (NESS), in which transport of
charge or heat through the system is 
possible. Instead, $\tilde{\rho}$  is a first-order
approximation of the thermal equilibrium state $\tilde{\rho}_{th} =
e^{-\beta H}/\tilde{Z}$ of the full Hamiltonian $H=H_0+V$. 
Normally (for example if invariance to time reversal
symmetry is not broken), no currents are generated in a thermal
equilibrium state  and therefore no steady-state
transport through the finite system can be described by this
approach~\cite{note}. 

To prove the above statement relating $\tilde{\rho}$  to $\tilde{\rho}_{th}$,
we take $\left<m\left|V\right|n\right>=0$ if
$\epsilon_{m}=\epsilon_{n}$, since as already discussed, the ``diagonal''
part $V_0$ of $V$  does not contribute to
transport. 
Consider then the eigenbasis $H| \tilde{n}\rangle =
\tilde{\epsilon}_n |\tilde{n}\rangle$, to first order perturbation
in $V$. Since $\left<m\left|V\right|n\right>=0$ for all
$\epsilon_{m}=\epsilon_{n}$,  we can apply the  first order
perturbation theory for non-degenerate states to all the states,
whether degenerate or not, to find
$\tilde{\epsilon}_n = \epsilon_n + {\cal O}(V^2)$ and $|\tilde{n}\rangle
= |n\rangle + \sum_{m, \epsilon_m\ne \epsilon_n}^{} \frac{\langle
  m|V|n\rangle}{\epsilon_n-\epsilon_m} |m\rangle+ {\cal O}(V^2)$. This
immediately leads to $\tilde{\rho}_{th}= \sum_{n}^{} {1\over \tilde{Z}}
e^{-\beta \tilde{\epsilon}_n} |\tilde{n}\rangle \langle
\tilde{n}| = \rho_0 + \delta \rho + {\cal O}(V^2)$, where
$\delta \rho$ is indeed given by Eq. (\ref{KuboI}). 

Therefore, we conclude that in order to describe an NESS with current
flow, one needs to go beyond viewing the finite
size system of interest as a closed system, and to explicitly consider
its connection to leads. We note that the effects of boundary
conditions  have been considered previously, {\em eg} in Refs.
\cite{Lebowitz, Dhar}. In particular, Ref. ~\cite{Dhar} derived a
Kubo-like formula that takes into consideration
cross-boundary currents in a stochastic approach.

In this Letter we present an alternative deterministic formulation
that explicitly 
considers the effects of coupling to leads (for charge transport) or
thermal baths 
(for heat transport) on the state of the system. It is based on the
Redfield equation \cite{Redfield} which describes the evolution of the
projected density matrix for the central system of interest, obtained
if we start from the Liouville-von Neumann equation for the total density
matrix describing the system+leads/baths and use the projection technique
\cite{KuboBook} to trace out the leads/baths. 

For simplicity, we assume coupling to thermal baths kept at
temperatures $T_{L/R}= T\pm\frac{\Delta T}{2}$
and investigate the heat transport in the resulting steady state. If
$\Delta T \ll T$, this leads to a Kubo-like formula which replaces
Eq. (\ref{Kubo2}). This approach can be generalized
straightforwardly to derive a Kubo-like formula for charge transport. 

The Redfield equation  has the general
form~\cite{KuboBook, Saito, Jinshan}:
\begin{equation}
\frac{\partial \rho(t)}{\partial t} = \left[ L_{H}+ L_L(T_L) +
  L_R(T_R)\right]\rho(t), \EqLabel{EEOM0}
\end{equation}
where $ L_{H}\rho=-i[H, \rho]$, just like for an isolated system, while
$L_{L/R}$ are additional terms that describe the effects of the
left/right thermal baths (assumed to be in equilibrium at their
corresponding temperatures $T_{L/R}$) on the evolution of the
system.  The expressions for $L_{L/R}$ depend on the Hamiltonian $H$ of
the system and on its coupling to
the baths (an example is provided below).

If $\Delta T \ll T$, we can Taylor expand $L_{L/R}$ and re-arrange the
Redfield equation to read:
\begin{equation}
\frac{\partial \rho(t)}{\partial t} = \left[L_{H_0} + L_B(T) +
  L_P(\Delta T)\right] \rho(t) = L\rho(t) , \EqLabel{EEOM}
\end{equation}
where $L_B(T) = L_R(T)+L_L(T)$ is the contribution from the thermal
baths if both are kept at the same temperature, while $L_P(\Delta T)$
collects the terms proportional to $\Delta T$.  Here we assume
 that $H=H_0$, {\em i.e.} that the thermal coupling 
does not induce an interaction $V$ in the system. For charge transport
such a term appears, and its Liouvillian $L_V$ should be grouped
together with $L_P$. 

Again, we are
interested in the $t\rightarrow \infty$, stationary state solution
$\rho$ of the above equation, {\em i.e.}
\begin{align}
L\rho=0 \EqLabel{openEx1}
\end{align}
which we assume to be unique for any value of $\Delta T$.  This means
that $L$ has a single zero eigenvalue, and all its other (transient)
eigenvalues have a negative real part.  Note that $L|_{\Delta T =0}
=L_{H_{0}}+ L_B(T)$ has this property. In fact, one can show that in
this case the $t \rightarrow \infty$ solution converges to the
expected thermal equilibrium for the system held at temperature $T$,
$\rho_0 = {1\over Z} e^{- \beta H_0}$~\cite{KuboBook}.

Eq. (\ref{openEx1}) can be solved numerically to find this
eigenstate. We call this solution 
$\rho_{ex}$. However, our goal is to obtain a Kubo-like formula. This
can be done by analogy with the calculation for the closed system
discussed in the beginning of this work. The first step is to separate
the Liouvillian $L$ of Eq. (\ref{EEOM}) into a ``large'' plus a
``small'' part. There are 
two possible choices: either 
take the ``large'' part to be $L^{(1)}_0 = L_{H_0} + L_B(T)$ with the
perturbation 
$\Delta L^{(1)} = L_P(\Delta T)$, or take $L^{(2)}_0 = L_{H_0}$ and let
$\Delta L^{(2)}= 
L_B(T)+ L_P(\Delta T)$ be the perturbation. 

We begin with the first choice.  $L^{(1)}_{0}=L_{H_0} + L_B(T)$ has
eigenvalues $\left\{L^{(1)}_{0,\mu}\right\}$ and left/right eigenvectors
$\left\{\left|\mathcal{L}_{\mu}\right)\right\}$, $
\left\{\left|\mathcal{R}_{\mu}\right)\right\}$.  As discussed, the
unique steady-state solution of $L^{(1)}_0 \rho_0=0$
is  $\rho_0={1\over Z} e^{-
  \beta H_0}$. The deviation 
$\delta \rho^{(1)}_{K}$ due to the perturbation 
$\Delta L^{(1)}$ is obtained like in Eq. \eqref{LRSsolution}:
\begin{align}
\delta\rho^{(1)}_{K}= \sum_{\mu}\int_{0}^{\infty}dt
e^{L^{(1)}_{0, \mu}t-\eta 
t}\left|\mathcal{R}_{\mu}\right)\left(\mathcal{L}_{\mu}\right|\Delta
L^{(1)} \rho_0 \notag \\ =
-\sum_{\mu}\frac{\left|\mathcal{R}_{\mu}\right)
\left(\mathcal{L}_{\mu}\right|}{L^{(1)}_{0, \mu}-\eta}\Delta
L^{(1)} \rho_0 =
-\sum_{\mu>0}\frac{\left|\mathcal{R}_{\mu}\right)
\left(\mathcal{L}_{\mu}\right|}{L^{(1)}_{0, \mu}}\Delta 
L^{(1)} \rho_0. \EqLabel{openLRS1}
\end{align} 
Note that the only divergent term, due to
$L^{(1)}_{0,0}=0$, disappears because
$\left(\mathcal{L}_{0}\right|\Delta L^{(1)} 
\rho_0=\left(\rho_0\right|\Delta
L^{(1)} \rho_0=0$.  To see why, we start from 
Eq. (\ref{openEx1}),  $L(\rho_0 + \delta \rho)=0$, project it on
$\left(\rho_{0}\right|$ and keep terms only to first order,
to find $0 =\left(\rho_{0}\right|\left(L^{(1)}_{0}+\Delta
L^{(1)}\right)\left(\rho_0 +\delta\rho\right) =
\left(\rho_{0}\right|\Delta L^{(1)} 
\rho_0$ since $L^{(1)}_0 \rho_0=0$.  As a result, 
Eq. (\ref{openLRS1}) has only regular contributions.  We denote
$\rho_0+ \delta \rho^{(1)}_{K}= \rho_{K}^{(1)}$.  A similar approach
using the eigenvalues and eigenvectors of
$L_{H_0} + L_B(T)$ has  been suggested in Ref. \cite{openkubo},
but for the
Lindblad equation~\cite{Lindblad} instead of the Redfield equation.

However, Eq. (\ref{openLRS1}) is difficult to use in practice;
finding all eigenstates of $L^{(1)}_0$ is a hard task unless the system 
has an extremely small Hilbert space. A computationally simpler
solution is obtained if we combine the eigenequation $L \rho =0$
with the constraint $Tr \rho=1$ into a regular system of coupled
equations $\bar{L} \bar{\rho} = \nu$, where, in matrix terms,
$\bar{L}$ is defined by replacing the first row of the equation $L\rho =0$ by
$Tr\left(\rho\right)=1$, so that $\nu$ is a vector whose
first element is $1$, all remaining ones being $0$. As a result
$\det\left(\bar{L}\right)\neq0$ while $\det\left(L\right)=0$. If solved
numerically, $\bar{L} \bar{\rho} = \nu$ produces the expected
solution $\rho_{ex}$. 

We can also solve it to obtain a Kubo-like formula by dividing
$\bar{L} = \bar{L}^{(1)}_0 + \Delta \bar{L}^{(1)}$. Again, the overbar
shows that in matrix terms, 
$\bar{L}^{(1)}_0$ is obtained from $L^{(1)}_0$ by replacing its first row with
$Tr \rho =1$, while $\Delta \bar{L}^{(1)}$ is obtained from $\Delta
L^{(1)}$ by replacing its first row with zeros. Then:
\begin{align}
\delta {\bar{\rho}}^{(1)}_{K} = - [\bar{L}^{(1)}_{0}]^{-1} \Delta
\bar{L}^{(1)}\rho_0. \EqLabel{openLRS2}
\end{align}
This is much more convenient because inverting the non-singular matrix
$\bar{L}_{0}^{(1)}$ is a much simpler task than finding
all the eigenvalues and eigenvectors of $L_{0}^{(1)}$. We have verified that
both schemes produce identical results. 

The second option is to take  $L^{(2)}_{0} = L_{H_0}$ and $\Delta L^{(2)} =
L_B(T)+L_P(\Delta T)$. In this case, we can still {\em choose} the
stationary solution associated with $L^{(2)}_0$ to be the thermal
equilibrium state at $T$, $\rho_0={1\over Z} e^{-\beta H_0}$. However,
this solution is no longer unique, in particular the thermal
equilibrium state corresponding to any other temperature 
satisfies $L_0^{(2)} \rho_{0}=0$. Expanding the analogue of Eq. (\ref{Kubo0})
in the eigenbasis of $H_0$, we  find:  
\begin{equation}
\delta\rho^{(2)}_{K}=-i
\sum_{n,m} \frac{\langle m|\Delta L^{(2)} \rho_0|n\rangle }{\epsilon_{m}-
  \epsilon_{n}-i\eta}|m\rangle \langle n|. 
\label{eq2}
\end{equation}
We call $\rho_{0} +
\delta\rho^{(2)}_{K}=\rho_{K}^{(2)}$. Note that unlike 
$\rho_{K}^{(1)}$ 
of Eq. (\ref{openLRS1}), this solution  has divergent
contributions from  states with $\epsilon_n=\epsilon_m$. As such,
it is analogous to the  Kubo formulae for infinite systems.

This similarity is not accidental. Kubo formulae for infinite
systems always ignore the coupling to the leads, taking $L_{0} =
L_{H_{0}}$ and {\em assuming} that $\rho_0=e^{-\beta H_0}/Z$, where
the temperature is arbitrarily chosen. Moreover, the driving
force leading to transport is not a term $L_P(\Delta T)$, since the
leads are ignored, but rather the addition of some potential $V$ to
$H_0$ leading to  $\Delta L = L_V$. Using only such a $V$ is rather questionable even
for charge transport, because of the previously mentioned problems
with the periodic boundary conditions, which may be negligible for
infinite size systems but are certainly not for finite-size
systems. For heat transport, using a $V$ 
to describe a variation in the gravitational potential~\cite{Luttinger} is
rather contrived, besides having the same boundary conditions
issues. Nevertheless, with these assumptions, Eq. (\ref{eq2}) maps
into the usual Kubo formula for an infinite system.  Thus, we can
think of Eq. (\ref{eq2}) with $L_P(\Delta T)$ included in $\Delta L$
as being a finite-size analog of the usual infinite-size Kubo formulae.
(As discussed, for a finite size system $L_P(\Delta T)$ cannot be
entirely replaced by an $L_V$, if steady-state transport is to be
established).

To see which of these two solutions --  the regular
solution $\rho_{K}^{(1)}$
or the singular solution  $\rho_{K}^{(2)}$ which is analogous
to formulae for infinite systems --  is the proper
one, we compare them against the exact numerical solution
$\rho_{ex}$ of Eq. (\ref{openEx1}) in the limit $\Delta T \ll T$.

We do this for a chain of $N$ spins $\frac{1}{2}$ coupled  by
nearest-neighbour exchange and placed in a 
magnetic field:
\begin{equation}
{\cal H}_0 = \sum_{i=1}^{N-1} J\vec{s}_{i}\cdot \vec{s}_{i+1}- B_z
\sum_{i=1}^{N}s_{i}^z
\nonumber
\end{equation}
while the heat baths are collections of bosonic modes,
\begin{align}
{\cal H}_B = \sum_{k, \alpha} \omega_{k,\alpha}
b^{\dag}_{k,\alpha}b_{k,\alpha}
\nonumber
\end{align}
where $\alpha=R, L$ indexes the right/left-side baths.  The
system-baths coupling 
is chosen as:
\begin{equation}
V_{int} = \lambda \sum_{k,\alpha} V^{(\alpha)}_{k}s_{i_\alpha}^{y}
\left(b^{\dag}_{k,\alpha} + b_{k,\alpha}\right)
\nonumber
\end{equation}
where $i_L =1$, $i_R=N$, {\em i.e.} the left/right bath is coupled to
the first/last spin and can induce its spin-flipping.

Using the projector technique~\cite{Saito, Jinshan}, the
equation of evolution for $\rho(t) = Tr_B \rho_t(t)$, where $\rho_t$
is the total density of states for the system+baths,  is found to
second-order perturbation theory in $V_{int}$, to be:
\begin{equation}
\EqLabel{evol} \frac{\partial \rho(t)}{\partial t} =-i[{\cal H}_0,
\rho(t)]-\lambda^2\sum_{\alpha=L,R}^{}\left( \left[s_{i_\alpha}^{y},
\hat{m}_{\alpha} \rho(t)\right] + h.c.\right)
\end{equation}
where $ \hat{m}_\alpha = s_{i_\alpha}^{y} \cdot \hat{\Sigma}_{\alpha}
$.  Here, $(\cdot )$ refers to the element-wise product of two
matrices, $\langle n| \hat{a}\cdot \hat{b} |m\rangle = \langle n|
\hat{a} |m\rangle \langle n|\hat{b} |m\rangle$. The bath matrices
$\hat{\Sigma}_{L,R}$ are defined in terms of the eigenstates of the
system's Hamiltonian ${\cal H}_0|n\rangle = \epsilon_n |n\rangle$ as:
\begin{eqnarray}
\nonumber \hat{\Sigma}_\alpha = \pi \sum_{m,n}^{} |m\rangle\langle n|
\left[\Theta\left(\Omega_{mn}\right)n_\alpha\left(\Omega_{mn}\right)
D_\alpha\left(\Omega_{mn}\right)|V^{(\alpha)}_{k_{mn}}|^2 \right. &&\\
\nonumber \left.  + \Theta\left(\Omega_{nm}\right)\left( 1+
n_\alpha\left(\Omega_{nm}\right)\right)
D_\alpha\left(\Omega_{nm}\right)|V^{(\alpha)}_{k_{nm}}|^2 \right] &&
\end{eqnarray}
where $\Omega_{mn} = \epsilon_{m}-\epsilon_{n}=-\Omega_{nm} $ and $k_{mn}$ is
defined by $\omega_{k_{mn},\alpha} = \Omega_{mn}$, i.e. is a bath mode
resonant with this transition. Furthermore, $\Theta(x)$ is the
Heaviside function, $n_\alpha(\Omega)=\left[e^{\beta_\alpha \Omega}
-1\right]^{-1}$ is the Bose-Einstein equilibrium distribution for the
bosonic modes of energy $\Omega$ at the bath temperature $T_\alpha
=1/\beta_\alpha$, and $D_\alpha(\Omega)$ is the bath's density of
states. The product
$D_\alpha\left(\Omega_{mn}\right)|V^{(\alpha)}_{k_{mn}}|^2$ is the
bath's spectral density function. For simplicity, we take it to be
a constant independent of $m$ and $n$. 

\begin{figure}[t]
\includegraphics[width=0.89\columnwidth]{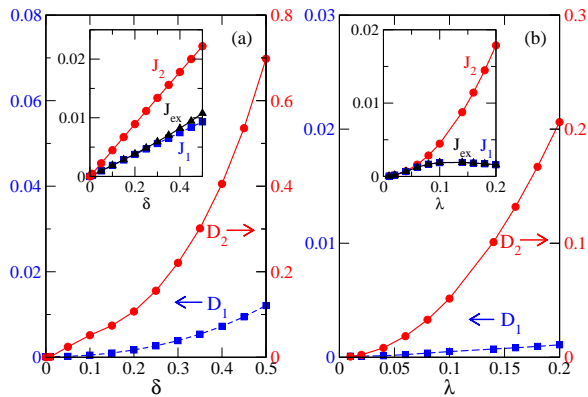}
\caption{\label{fig1} (color online) $D_1$ (squares) and $D_2$ (circles) of
Eq. (\ref{norm}) (a) vs. $\delta=\Delta T/2T$ at a fixed $\lambda = 0.1$, 
and (b) vs. $\lambda$ at a fixed $\delta = 0.1$, 
for $N=8, J=0.1, B_z=1$. The insets show the steady-state   thermal
current calculated with $\rho_{ex}$ 
(triangles) and $\rho_{K}^{(1,2)}$ (squares and circles).
See text for more details.} 
\label{N=8}
\end{figure}

Eq. (\ref{evol}) is thus a particular example of the general
Eq. (\ref{EEOM0}), with the last two terms coming from the coupling to
the two thermal baths. Since the temperatures $T_{L/R}$ enter only in
the Bose-Einstein occupation numbers, it is straightforward to expand
them when $T_{L/R}= T\pm {\Delta T\over 2}, \Delta T \ll T $ and so to
identify $L_B(T)$ and $L_P(\Delta T)$. 

We characterize the distance between the exact numerical solution
$\rho_{ex}$ and the two possible Kubo solutions $\rho_{K}^{(i)}$,
$i=1,2$ by calculating the norm:
\begin{align}
\label{norm}
D_{i} = \sqrt{\sum_{n,m}\left|\langle
  n|\rho_{ex}-\rho_{K}^{(i)}|m\rangle\right|^2}.
\end{align}
For the proper solution, this difference should be small but finite
due to higher-order perturbation terms. 

Results typical of those found in all the cases we investigated are
shown in Fig. \ref{fig1} for $N=8, J=0.1, B_{z}=1.0$. In Fig.
\ref{fig1}(a), we plot $D_{1,2}$ vs. $\delta=\Delta T/ 2T$ for a fixed
system-baths coupling $\lambda = 0.1$, while in Fig. \ref{fig1}(b) we
show them vs. $\lambda$, for $\delta T = 0.1$.  In both cases, $D_2$
(circles, axis on the right) is very large. In fact, because of the
singular contributions from $\epsilon_n=\epsilon_m$ states, $D_2$ is
divergent, its magnitude being controlled by the cutoff $\eta$ used
($\eta = 10^{-5}$ here). In contrast, $D_1$ (squares, left axis) has a
small value independent of $\eta\rightarrow 0$, showing that the
regular $\rho^{(1)}_{K}$ is the proper Kubo solution. The insets show
the thermal current calculated with $\rho_{ex}$, $\rho_K^{(1)}$ and
$\rho_K^{(2)}$ (triangles, squares, respectively circles). $J_2$
becomes independent of $\eta$ as $\eta\rightarrow 0$, however, unlike
$J_1$ it is quite different from the exact solution. This again
confirms that $\rho^{(1)}_{K}$ is the proper Kubo solution.

Fig. \ref{fig1}(b) also shows that the answer depends on the details
of the coupling to the baths. This is not surprising for a finite-size
system: the intrinsic conductance of the system is added to comparable
``contact'' contributions from the interfaces between the
system and the baths, and experiments measure the total conductance. It follows
that quantitative modeling of transport in finite systems will 
require a careful consideration of the entire experimental set-up.

To conclude, we make two claims regarding the proper Kubo
formulae to be used for finite-size systems. The first is that the
coupling to leads/baths simply cannot be ignored, as is usually done
for infinite-size systems. Instead, it has to be considered explicitly
if steady-state transport is to be established in the
system. Secondly, we showed that the proper way to derive a Kubo-like
formula in such cases leads to a well-behaved result. This is to be
contrasted with the formulae typically used in literature, similar to
those valid for infinite-size systems, and which have singular
contributions from eigenstates with $\epsilon_n= \epsilon_m$. In
infinite systems this is not a problem because the spectrum is
continuous and the final result is still well-behaved. However, for
finite-size systems the spectrum is discrete and the singularities
cannot be avoided. This is clearly unphysical: a finite-size system
cannot have singular response functions. Of course, in reality the
``eigenstates'' of the open system are no longer
sharp, instead they acquire a finite lifetime due to tunneling of
charge/heat into and out of the leads/baths. The need to properly consider
the effects of the leads/baths on the system  is therefore unavoidable.

{\bf Acknowledgements:} We thank Ian Affleck for many discussions and
suggestions. This work was supported by NSERC and
CIFAR.


\begin{thebibliography}{99}

\bibitem{Kubo} R. Kubo, J. Phys. Soc. Japan, {\bf
12}, 570-586(1957).

\bibitem{Luttinger} J.M. Luttinger, Phys. Rev. {\bf 135}, A1505(1964).

\bibitem{Heidrich} F. Heidrich-Meisner {\em et al.}, Phys. Rev. B{\bf
71}, 184415 (2005), and
references therein.

\bibitem{com1} Even this is rather subtle, as can be seen
using eigenstates $c_k^{\dagger}|0\rangle$ for a non-interacting
system with  $V_0=0$.

\bibitem{note1} The fact that one should not sum over 
$\epsilon_n=\epsilon_m$ states in the Kubo formula for finite systems
can also be verified explicitly for toy models, {\em eg.}
$H_{0}=\frac{1}{2}\left(p^2+x^2\right)$ and $ 
V=\frac{\lambda}{2}x^2$, which are exactly solvable.  
It is easy to check that Eq. \eqref{KuboI} gives the correct answer to
${\cal O}(\lambda^2)$, while inclusion of $\epsilon_{m}=\epsilon_{n}$
terms gives  wrong results. 

\bibitem{note} Non-equilibrium stationary states allowing the transport
of charge currents are possible in finite size systems with periodic
boundary condition, if a time-dependent magnetic flux is threaded
through. However, this is not how such 
measurements are  performed in practice. 

\bibitem{Lebowitz} J. L. Lebowitz and H. Spohn, J. Stat. Phys. {\bf
95} 333(1999).
\bibitem{Dhar} A. Kundu {\em et al.}, %, A. Dhar and O. Narayan,
J. Stat. Mech. L03001(2009).

\bibitem{Redfield} K. Blum, {\it{Density Matrix Theory and
Applications: Physics of Atoms and Molecules}}, (Plenum, 1996).

\bibitem{KuboBook} R. Kubo, M. Toda and N. Hashitsume, {\it
Statistical Physics II: Nonequilibrium Statistical Mechanics}
(Springer, 1998).

\bibitem{Saito} K. Saito {\em et al.}, Phys. Rev. E, {\bf 61},
2397(2000).

\bibitem{Jinshan} J. Wu and M. Berciu, arXiv:1003.1559.

\bibitem{openkubo} M. Michel {\em et al.}, %, J. Gemmer and G. Mahler,
Eur. Phys. J. B, {\bf 42}, 555 (2004).


\bibitem{Lindblad} G. Lindblad, Commun. Math. Phys. {\bf{48}},
119(1976).

\end{thebibliography}
\end{document}